\documentclass[prl,superscriptaddress,twocolumn,showpacs,amsmath,amssymb]{revtex4-1}

\usepackage{graphicx}
\usepackage{bm}
\usepackage[breaklinks=true,colorlinks,citecolor=blue,linkcolor=blue,urlcolor=blue]{hyperref}
\usepackage{color}
\usepackage{amsmath}
\usepackage{datetime}
\usepackage[normalem]{ulem}

\preprint{Submission target: Nano Letters -- \today { }-- \currenttime}

\begin{document}

\title{InAs nanowire superconducting tunnel junctions: spectroscopy,\\ thermometry and nanorefrigeration}

\author{Jaakko Mastom\"aki}
\affiliation{NEST, Scuola Normale Superiore and Istituto Nanoscienze-CNR, Piazza S. Silvestro 12, I-56127 Pisa, Italy}
\affiliation{Nanoscience Center, Department of Physics, University of Jyvaskyla, 40014 Jyv\"askyl\"a, Finland}
\author{Stefano Roddaro}
\affiliation{NEST, Scuola Normale Superiore and Istituto Nanoscienze-CNR, Piazza S. Silvestro 12, I-56127 Pisa, Italy}
\author{Mirko Rocci}
\affiliation{NEST, Scuola Normale Superiore and Istituto Nanoscienze-CNR, Piazza S. Silvestro 12, I-56127 Pisa, Italy}
\author{Valentina Zannier}
\affiliation{NEST, Scuola Normale Superiore and Istituto Nanoscienze-CNR, Piazza S. Silvestro 12, I-56127 Pisa, Italy}
\author{Daniele Ercolani}
\affiliation{NEST, Scuola Normale Superiore and Istituto Nanoscienze-CNR, Piazza S. Silvestro 12, I-56127 Pisa, Italy}
\author{Lucia Sorba}
\affiliation{NEST, Scuola Normale Superiore and Istituto Nanoscienze-CNR, Piazza S. Silvestro 12, I-56127 Pisa, Italy}
\author{Ilari J. Maasilta}
\affiliation{Nanoscience Center, Department of Physics, University of Jyvaskyla, 40014 Jyv\"askyl\"a, Finland}
\author{Nadia Ligato}
\affiliation{NEST, Scuola Normale Superiore and Istituto Nanoscienze-CNR, Piazza S. Silvestro 12, I-56127 Pisa, Italy}
\author{Antonio Fornieri}
\affiliation{NEST, Scuola Normale Superiore and Istituto Nanoscienze-CNR, Piazza S. Silvestro 12, I-56127 Pisa, Italy}
\author{Elia Strambini}
\affiliation{NEST, Scuola Normale Superiore and Istituto Nanoscienze-CNR, Piazza S. Silvestro 12, I-56127 Pisa, Italy}
\author{Francesco Giazotto}
\affiliation{NEST, Scuola Normale Superiore and Istituto Nanoscienze-CNR, Piazza S. Silvestro 12, I-56127 Pisa, Italy}

\begin{abstract}

We demonstrate an original method -- based on controlled oxidation -- to create high-quality tunnel junctions between superconducting Al reservoirs and InAs semiconductor nanowires. 
We show clean tunnel characteristics with a current suppression by over $4$ orders of magnitude for a junction bias well below the Al gap $\Delta_0 \approx 200\,\mu {\rm eV}$. 
The experimental data are in close agreement with the BCS theoretical expectations of a superconducting tunnel junction. The studied devices combine small-scale tunnel contacts working as thermometers as well as larger electrodes that provide a proof-of-principle active {\em cooling} of the electron distribution in the nanowire. A peak refrigeration of about $\delta T = 10\,{\rm mK}$ is achieved at a bath temperature $T_{bath}\approx250-350\,{\rm mK}$ in our prototype devices. This method opens important perspectives for the investigation of thermoelectric effects in semiconductor nanostructures and for nanoscale refrigeration.
\end{abstract}

\maketitle

The control over the heat flow and the local electron distribution in a nanodevice represents a crucial experimental challenge~\cite{giazotto2012josephson,altimiras2010non,muhonen2012micrometre} with an important impact both on the solution of key open problems in fundamental physics and on development of future device applications~\cite{giazotto2006opportunities,fornieri2016towards}. In particular, the recent progress of thermoelectric physics in nanostructured materials offers fascinating new perspectives for the realization of more efficient solid-state heat pumps for energy conversion~\cite{dresselhaus2007new,wu2013large,vineis2010nanostructured} and/or for the creation of self-cooling nanodevices where the electron or the phonon system in the active region can be refrigerated below the phonon bath~\cite{giazotto2006opportunities,muhonen2012micrometre}. The progress in these fields calls for the development of novel methods to reliably control heat and measure the device thermoelectric parameters with a nanometer-scale precision~\cite{li2003thermal,yazji2015complete,roddaro2013giant,tikhonov2016local}. Local electronic cooling can be relevant for improving the device performance either in terms of noise, sensitivity or decoherence~\cite{giazotto2006opportunities} and for finding a role in advanced applications, including topological quantum computation~\cite{mourik2012signatures,plissard2013formation,chang2013tunneling,larsen2015semiconductor} and ultrasensitive radiation detection~\cite{miller2008high,giazotto2008ultrasensitive}. In addition, the manipulation of heat is at the basis of the emerging field of coherent caloritronics~\cite{giazotto2012josephson,martinez2014quantum,fornieri2015nanoscale,martinez2015rectification} and it could be crucial in solving important standing fundamental problems in condensed matter physics, including quantum thermodynamics and the study of the elusive Majorana fermions in solid-state systems~\cite{mourik2012signatures,leijnse2014thermoelectric,lopez2014thermoelectrical}.

\begin{figure}[ht!]
\begin{center}
\includegraphics[width=\columnwidth]{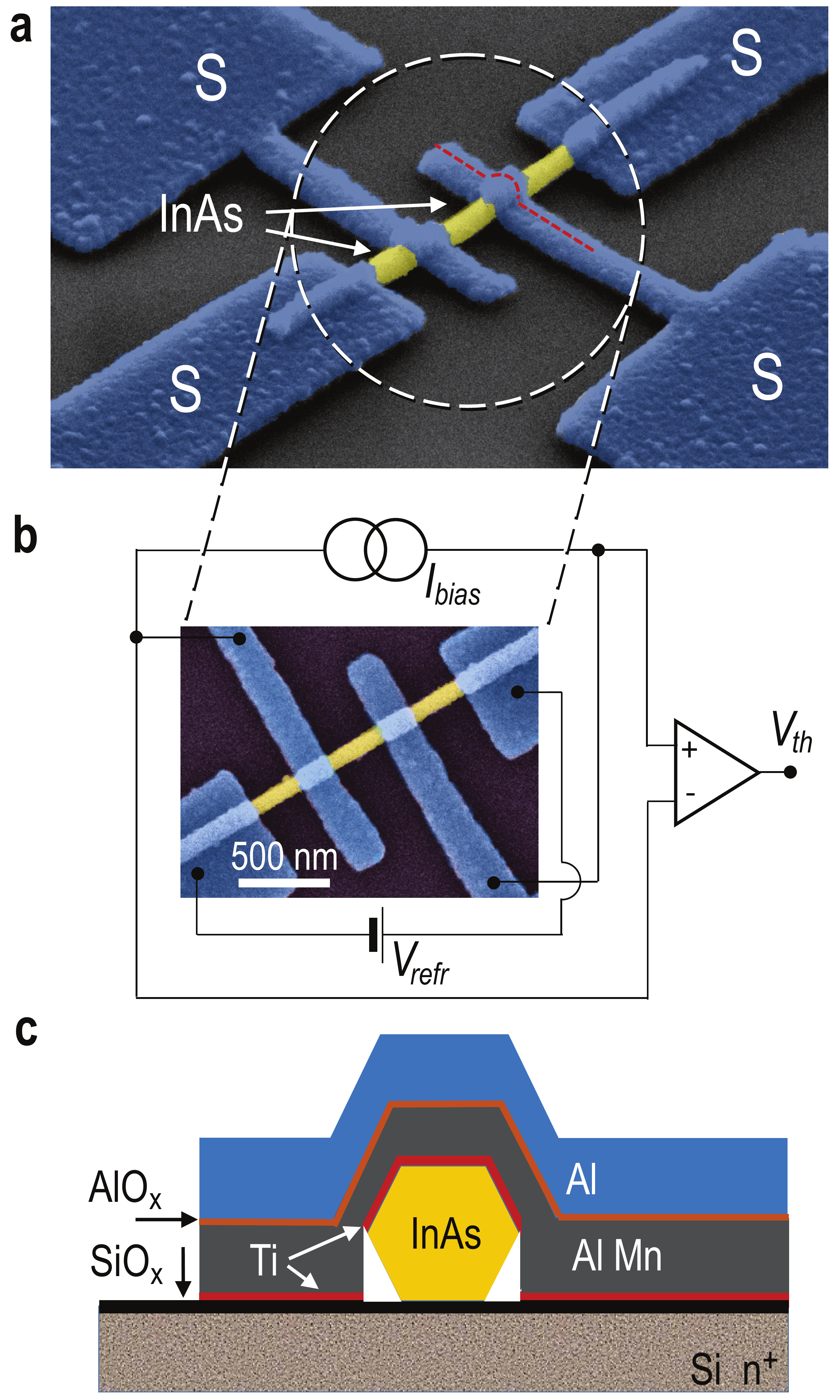}
\caption{{\bf Device architecture.} (a) Scanning electron micrograph of a typical device: four tunnel junctions are created between the superconductive electrodes (S) and an $n$-doped InAs nanowire. Two inner $200\,{\rm nm}$-wide contacts are used to measure the electron temperature in the InAs nanostructure. Two larger contacts are fabricated at the two ends of the nanowire and are used to extract hot carriers from the InAs and to refrigerate its electron system below the bath temperature. The sample was imaged at $50^\circ$ angle. (b) Local thermometry is achieved by biasing the inner contacts with a constant current $I_{bias}$ and by measuring the resulting voltage drop $V_{th}$ in four-wire scheme; refrigeration is achieved by biasing the outer contact with a voltage $V_{r\!e\!f\!r}$. (c) Cross-sectional view of the tunnel junction along the dashed red line in (a): the NIS barrier is obtained by a controlled {\em in-situ} oxidation of a non-superconductive Al layer containing Mn impurities with a thickness of $50\,{\rm nm}$. A conventional $50\,{\rm nm}$-thick superconductive Al layer is evaporated on top of the oxide barrier. A thin $5\,{\rm nm}$ layer of Ti is deposited between the NW and the AlMN film to promote adhesion.}
\label{fig:Cartoon}
\end{center}
\end{figure}

Hybrid architectures combining superconductive elements with normal metals represent a bright example of refined technology to locally measure and manipulate heat at low temperatures and have been the subject of an intense research effort~\cite{giazotto2006opportunities,fornieri2016towards,muhonen2012micrometre}. In particular, devices integrating normal-insulator-superconductor (NIS) tunnel junctions between a normal metal (N) and Al~\cite{pekola2004limitations} or other superconductors (S)~\cite{quaranta2011cooling,nevala2012sub} have yielded the demonstration of significant nanorefrigeration in the milliKelvin regime. A similar technology has been so far hard to achieve in the context of semiconductor nanostructures, given the notorious technical challenges that the realization of clean semiconductor interfaces entails~\cite{gunnarsson2015interfacial}. While alternative approaches exist~\cite{svensson2013nonlinear}, they cannot offer a comparable performance, in particular in terms of cooling power~\cite{pekola2004limitations}. In this Letter, we focus on combining superconductive tunnel contacts with the technology of self-assembled semiconductor nanowire (NW). NW technology has been recently used to create advanced nanodevices in the context of thermoelectrics~\cite{wu2013large}, Josephson devices~\cite{doh2005tunable,roddaro2011hot,giazotto2012josephson}, single-electron manipulation~\cite{bjork2002one,roddaro2011manipulation,romeo2012electrostatic,rossella2014nanoscale}, topological quantum computing and Majorana physics~\cite{mourik2012signatures,plissard2013formation,chang2013tunneling,larsen2015semiconductor}, but none of these applications utilized hard tunnel junctions. In this work, we exploit the controlled oxidation of a thin layer of Al alloy to fabricate junctions with a controlled transparency, and use InAs nanowires (NWs) to create SI(NW)IS tunnel devices. We demonstrate that this technique can be used to obtain state-of-the-art tunnel characteristics, which can be exploited to perform a sensitive local thermometry using nanoscale contacts. In addition, large junctions are used to demonstrate electron cooling of an individual NW: a peak refrigeration of $\delta T=10\,{\rm mK}$ is demonstrated in the current device architecture, with the expectation that it can be improved significantly with more optimal device geometry.

Devices are built starting from degenerately Se-doped $n$-type InAs NWs~\cite{viti2012se}, grown by chemical beam epitaxy (Methods). The wires are deposited on a SiO$_2$/Si substrate by dropcasting and contacted by metal electrodes fabricated by electron-beam lithography. The ${\rm SiO_2}$ layer is $300\,{\rm nm}$ thick and provides a good isolation between the device and the Si substrate. Four electrodes are fabricated on each NW, as visible in Fig.~1a: two small-scale, $200\,{\rm nm}$-wide fingers at the central part of the nanostructure, and two larger contacts at the two ends of the NW. The central electrodes are designed to work as tunnel thermometers and are separated by about $300\,{\rm nm}$. The outer tunnel electrodes have larger junction areas (about $90\times600\,{\rm nm^2}$) and are designed to control the electron distribution in the NW. Their nominal separation from the inner electrodes is also $300\,{\rm nm}$. A simplified sketch of the set-up used to measure the electronic temperature and to refrigerate the electrons in the NWs is shown in Fig.~1b. The key details of the contact technology discussed here are illustrated in Fig.~1c. Tunnel contacts are obtained by multiple steps of electron-beam lithography, evaporation and controlled {\em in-situ} oxidation. After the e-beam patterning and development, the InAs surface is first passivated using a ${\rm (NH_4)_2S_x}$ solution~\cite{suyatin2007sulfur} to remove the native oxide. The sample is then immediately transferred to the vacuum chamber of an e-beam evaporator where a $5/50\,{\rm nm}$-thick layer of ${\rm Ti/AlMn}$ is evaporated. The purpose of the Mn impurities in the alloy is to quench superconductivity in the Al-based layer~\cite{ruggiero2004dilute}, so that it can be used to create a thin normal-metal interlayer between the oxide barrier on top of it and the InAs crystal. In the current architecture, a residual superconductive effect can be expected due to the Ti layer. After the first evaporation, the sample is transferred to an oxidation chamber where the AlMn layer is exposed to a $0.2-0.4\,{\rm Torr}$ of ${\rm O_2}$ for $5$ minutes, which is expected to yield a $1-2\,{\rm nm}$-thick oxide layer. Importantly, Mn impurities are known to have no detrimental effects on the barrier quality~\cite{ruggiero2004dilute}. After the oxidation, a residual AlMn layer with a thickness of about $50\,{\rm nm}$ remains as a buffer between the oxide and the InAs. In principle, its thickness could be reduced by evaporating a thinner AlMn layer, but at the risk of non-homogeneity and thus of a lower-quality oxide barrier. Finally, a $50\,{\rm nm}$-thick layer of superconductive Al is added on top of the AlO$_x$. 

\begin{figure}[h!]
\begin{center}
\includegraphics[width=\columnwidth]{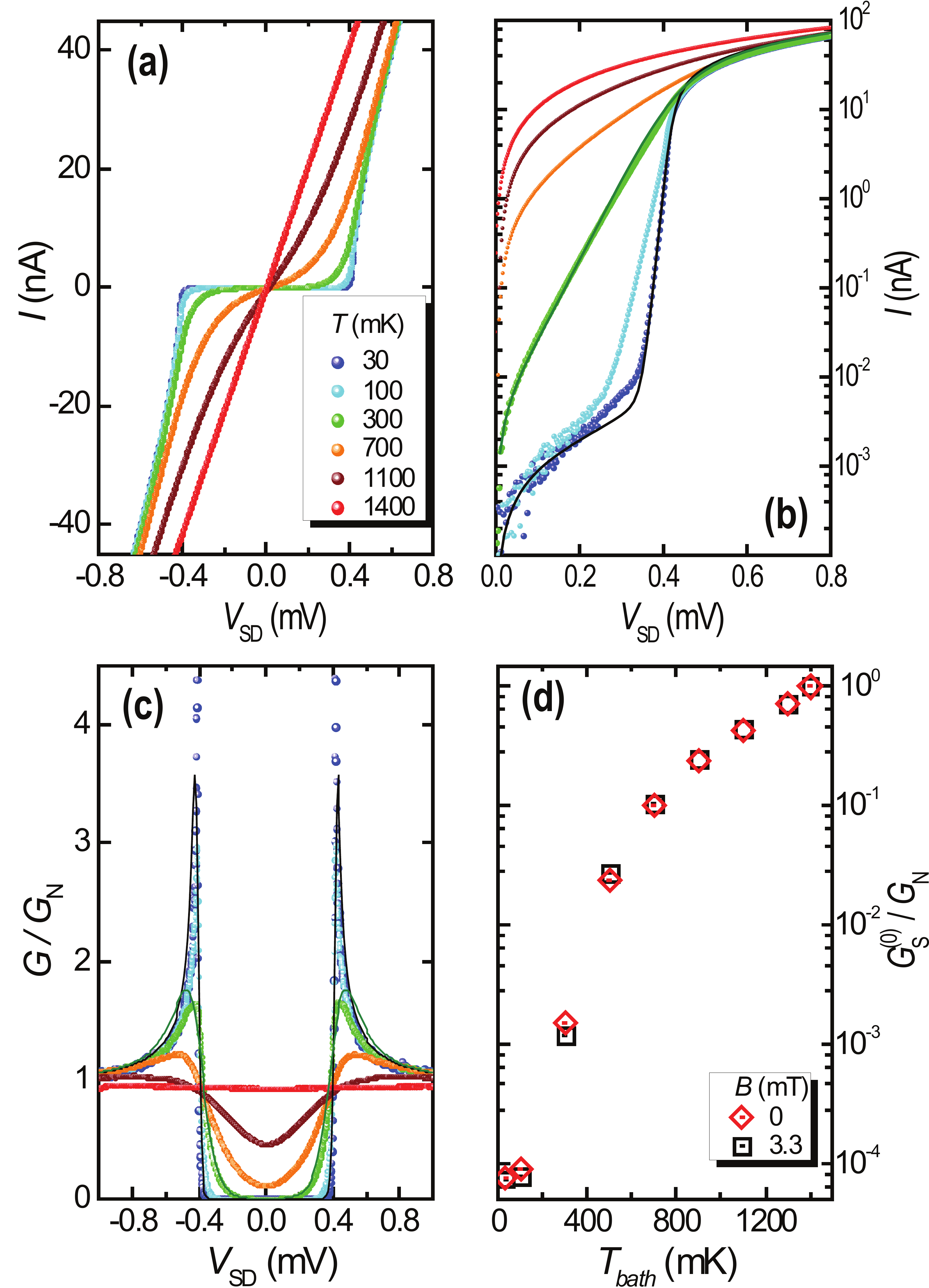}
\caption{{\bf Tunnel spectroscopy.} (a) Selected $IV$ curves for a typical SI(NW)IS configuration involving the inner device contacts, in the presence of a small out-of-plane magnetic field of $3.3\,{\rm mT}$. (b) Same in semi-log scale. At zero bias, the current $I$ is strongly quenched by the superconductive gap in the Al contacts, demonstrating the achievement of high-quality tunnel junctions. Selected experimental data (dots) are compared to BCS theory (solid lines), with good agreement. (c) As expected for a well-behaved tunnel junction, the differential conductance $G=dI/dV$ is proportional to the convolution of the density of states on the N and S sides of the junctions and the derivative of the Fermi-Dirac distribution. A gap of $\Delta_0 \approx 208\pm1\,{\rm meV}$ can be deduced from a theoretical fit of the experimental data using BCS theory. (d) Normalized zero-bias conductance vs bath temperature.}
\label{fig:Cartoon}
\end{center}
\end{figure}

\begin{figure}[ht!]
\begin{center}
\includegraphics[width=\columnwidth]{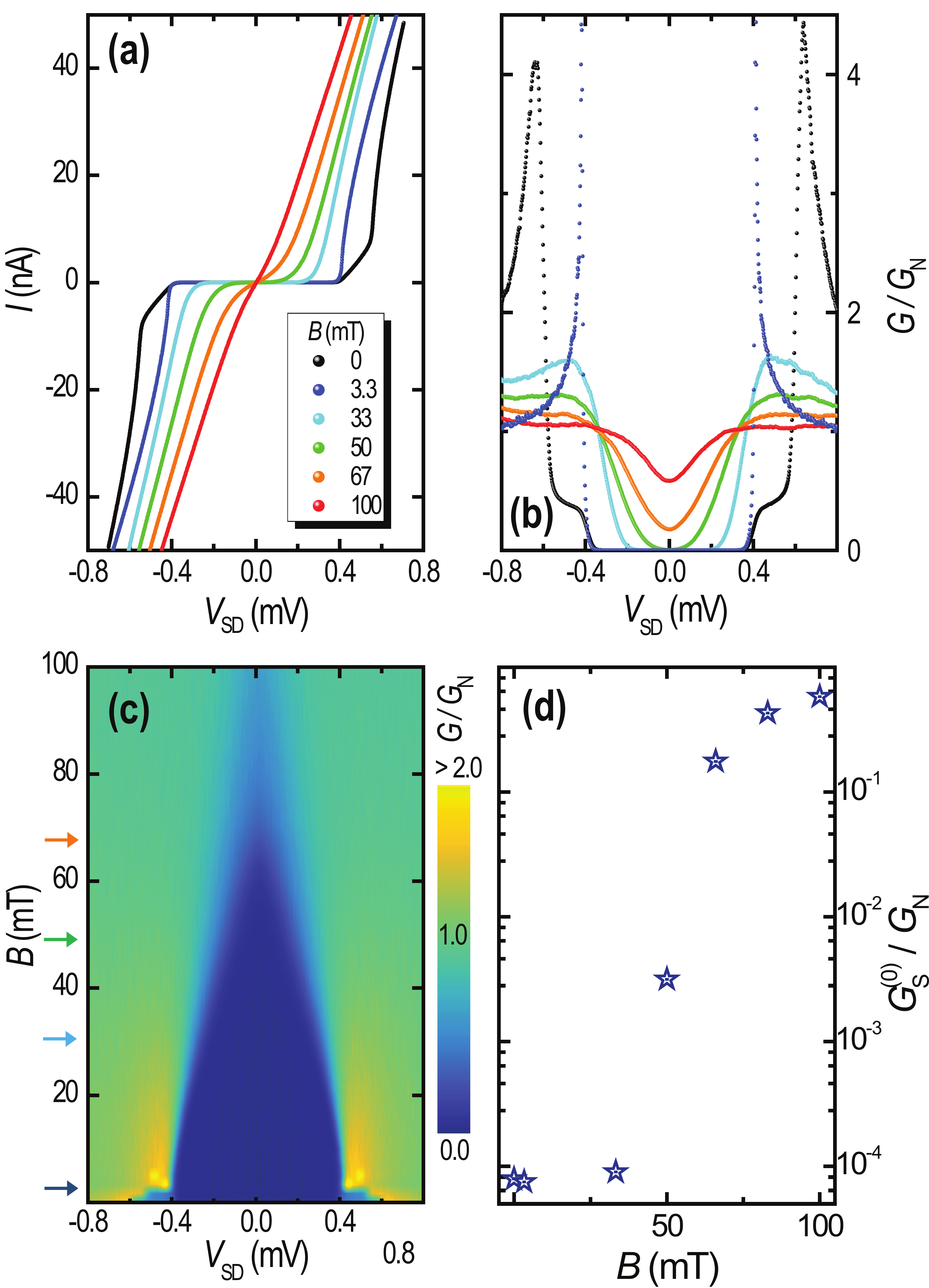}
\caption{{\bf Magnetic field suppression.} The junctions display a non-trivial behavior as a function of the magnetic field $B$, as visible from the $IV$ curves in panel (a) and from the corresponding differential conductance traces in panel (b). In particular, normalized differential conductance at $B=0$ highlights the presence of a double step in the density of states. The possible origin of this phenomenon is a residual pairing on the NW side of the tunnel junctions (see text). The full evolution of the effect is better visible in panel (c), reporting the conductance colorplot versus $B$. A small magnetic field ($\approx 3\,{\rm mT}$) quenches the residual proximity in the NW while the rest of the evolution can be understood as due to the quenching of superconductivity in the Al electrodes, as also visible from the evolution of zero-bias conductance in panel (b). The arrows in panel (c) indicate the intermediate values of magnetic field for the plots in panels (a) and (b). (d) The evolution of the normalized zero-bias differential conductance as a function of $B$ indicates a critical field of Al of about $100\,{\rm mT}$.}
\label{fig:Cartoon}
\end{center}
\end{figure}

\begin{figure}[ht!]
\begin{center}
\includegraphics[width=\columnwidth]{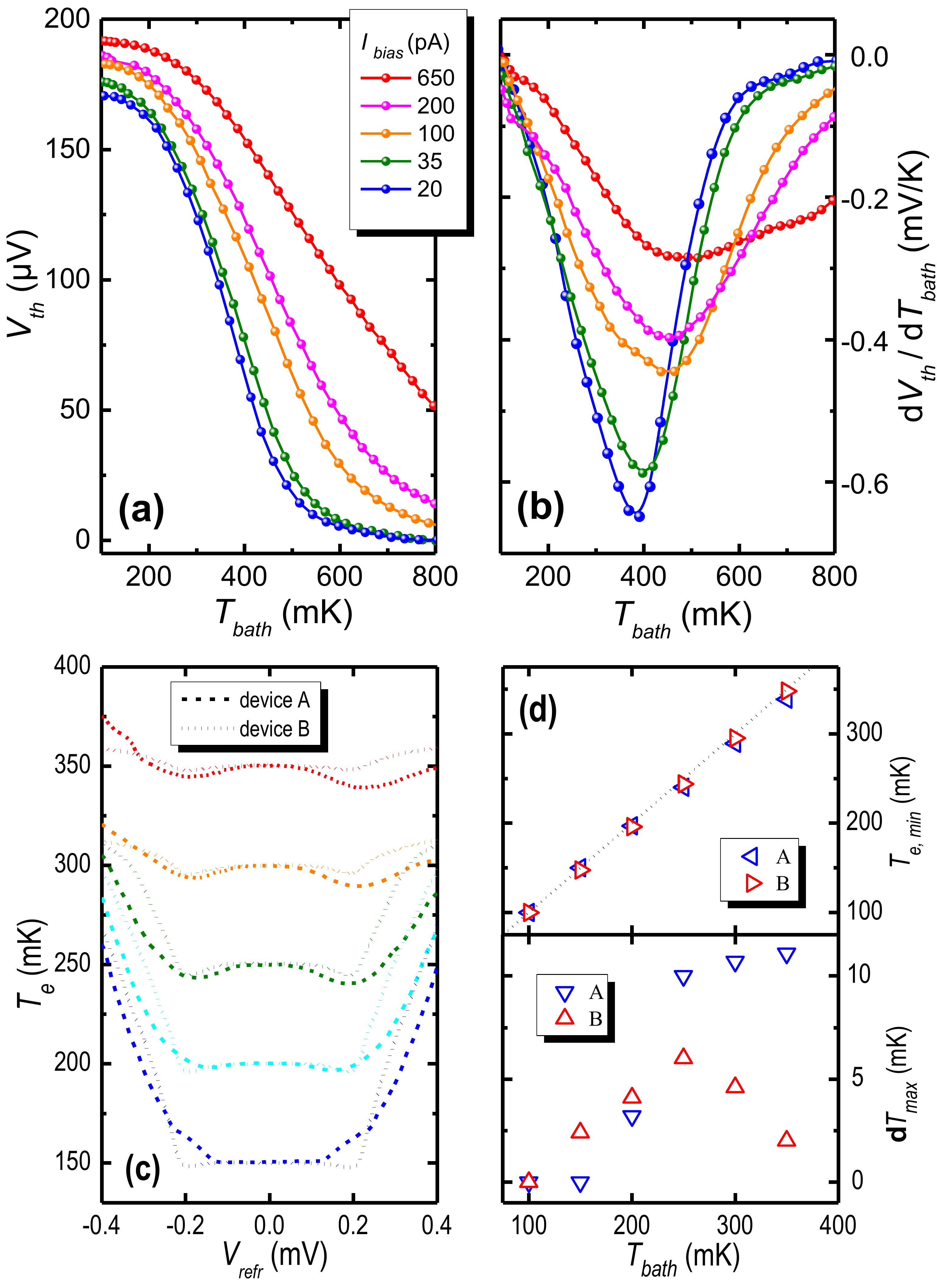}
\caption{{\bf Thermometry and cooling.} The inner SI(NW)IS junctions work as a sensitive electronic thermometer, as visible from the $V_{th}$ voltage response at fixed current bias $I_{bias}$ in panel (a). The thermometer response is even more evident from the $dV_{th}/dT_{bath}$ plot in panel (b), indicating a top responsivity of over $0.6\,{\rm mV/K}$. The outer and larger SI(NW)IS junction can be used to {\em tailor} the electron distribution in the NW. The plot in panel (c) shows the cooling of the electron temperature $T_e$ as a function of the refrigerator bias $V_{r\!e\!f\!r}$ measured with biasing the thermometer junctions with $ I_{bias} = 20\,{\rm pA} $. A top cooling of $\approx 10\,{\rm mK}$ could be achieved starting from $T_{bath}\approx 250-350\,{\rm mK}$, visible from temperature data extracted in panel (d). The dotted line in the upper panel shows the equilibrium $T_e = T_{bath}$. A and B indicate two different NW refrigerators}
\label{fig:Cartoon}
\end{center}
\end{figure}

Contacts obtained with the described procedure typically yielded a low-temperature normal-state resistance of about $6.5\,{\rm k\Omega}$ and $2.5\,{\rm k\Omega}$ for the small- and large-area NIS barriers, respectively. The NWs typically displayed an additional $\approx 2\,{\rm k\Omega}$ resistance between the two outer contacts. These resistances can be determined separately for each contact and the NW, by measuring every pair combination. The contacts exhibit almost ideal IV characteristics, as shown in Fig. 2, if operated in a small out-of-plane magnetic field $B=3.3\,{\rm mT}$. In these experimental conditions the $IV$ characteristics are consistent with what expected for a SINIS device where, in our case, the N portion of the device is implemented by the InAs NW plus two thin and strongly-coupled AlMn interlayers. The temperature evolution of the $IV$ curves is visible in Fig.~2a, proving the high quality of the tunnel junctions. In the semilog plot in Fig.~2b, a suppression of about $4$ orders of magnitude with respect to the normal state is observed in the current at the base temperature $T_{bath}=30\,{\rm mK}$, when the bias $V_{SD}$ is smaller than the superconductive gap $\Delta_0$. The quality of the tunnel junction can be even better appreciated from the normalized differential conductance $G/G_N$ plot of Fig.~2c, where $G=dI/dV$ and $G_N$ is the normal-state conductance evaluated at high bias voltages ($V_{SD} \gg \Delta_0$) and low temperatures. Experimental results (dots) in panels (b) and (c) are directly compared with BCS predictions (continuous lines), and a good agreement is obtained assuming $\Delta_0 = 208\pm 1\,\mu{\rm eV}, R_N = 9.4\,{\rm k\Omega} $ and Dynes parameter $\gamma = 8\cdot 10^{-5} $ (see Methods). The best fits for the data measured at $ 30\,{\rm mK} $ and $ 300\,{\rm mK} $ bath temperatures were obtained by setting electron temperatures as $ 53\,{\rm mK} $ and $ 275\,{\rm mK} $, respectively. In Fig.~2d we report the corresponding suppression of differential conductance at zero magnetic field and at a small off-plane $B=3.3\,{\rm mT}$. The field has a marginal  effect on the zero-voltage conductance but it is important to suppress extra features in the $IV$ which are likely caused by residual pairing effects discussed next.

The detailed evolution of the tunnel characteristics as a function of the out-of-plane magnetic field $B$ is visible in Fig.~3. $IV$ curves and the corresponding normalized differential conductance $G/G_N$ are shown in panels (a) and (b), respectively. The junction displays a complex behavior at zero field, with additional structures in $G$: the effect indicates  a residual superconductive gap on the NW-side of the tunnel junction, as a likely consequence of the presence of the Ti layer. The $IV$ anomalies are easily quenched by the application of a few milliTesla magnetic field. The full evolution of the normalized conductance as a function of $B$ can be better appreciated from the colorplot of Fig.~3c displaying also the expected quenching of superconductivity in the Al electrodes. The impact of the zero-field anomaly visible in panels (a) and (b) can also be easily identified in the colorplot at $B< 3 $mT. A critical field $B_c \gtrsim 100\,{\rm mT}$ is obtained from the colorplot, as also visible from the conductance plot $G(B)$ in panel (d). 

Superconducting tunnel contacts can be used both for measuring the electronic temperature and to actively cool the electron distribution in the NW \cite{giazotto2006opportunities}. The inner narrower contacts have a larger barrier resistance and are more suited for a local probing of the temperature. The typical response voltage $V_{t\!h}$ for one of the inner SI(NW)IS junctions is plotted in Fig.~4a, for different values of the current bias $I_{bias}$. A responsivity up to about $0.6\,{\rm mV/K}$ is obtained for a $20\,{\rm pA}$ bias (see Fig.~4b). The wider tunnel contacts fabricated at the end of the NWs were used to demonstrate local refrigeration of an individual NW. Two devices, named $A$ and $B$, were tested in such a cooling configuration: the corresponding steady-state temperatures for different bias voltages $V_{r\!e\!f\!r}$ are plotted in Fig.~4c. The electronic temperature ($T_e$) have been extracted from the thermometer calibration of Fig.~4a. The largest cooling is achieved for $|qV_{r\!e\!f\!r}|\approx \Delta_0$, while heating is obtained for larger bias, as expected for a single NIS cooler \cite{giazotto2006opportunities}. This behavior is consistent with a non-ideal behavior of the current devices, where typically one of the two cooling contacts is too transparent. The minimum electron temperatures $T_{e,min}$ for different bath temperatures are plotted in the upper panel of Fig.~4d. The top temperature reduction $\delta T_{max} = T_{bath}-T_{e,min}$ having its maximum value $~10\,{\rm mK}$ at $T_{bath}\approx 250-350\,{\rm mK}$ is shown in the lower panel. While the cooling effect remainMas modest, it demonstrates that our tunnel junction can be effectively used to reduce the electronic temperature in an individual NW. Various non-ideal factors hamper the device performance in the current architecture: (i) the tunnel barrier cools the whole AlMn region, which is wider that the NW (see sketch in Fig.~1c); (ii) the barrier opacity is still not optimal, and typically one of the two contacts of the refrigerator is too transparent. A possible route for achieving a better performance consists of patterning different geometries for the AlMn layer and for the Al layer, so to achieve tunnel junctions which are more controlled and have a smaller area in correspondence with the NW body.

In conclusion, we have demonstrated an original technique for the fabrication of superconducting tunnel junctions on InAs-based semiconductor NWs. The junctions have been shown to be suitable for low-temperature thermometry, and electronic cooling was demonstrated at optimal biasing conditions. The relatively small cooling observed in the present devices does not represent a fundamental limitation and can be largely improved by using an optimized contact geometry. This technology can have a large impact in cryogenic circuits requiring local cooling, and can benefit a large number of nanoscience fields, such as sensing, quantum computation and quantum technology in general.

{\bf Acknowledgments.} S.R., V.Z., L.S, D.E. and M.R. acknowledge the financial support by CNR, through the bilateral projects with RFBR (Russia), and by Scuola Normale Superiore. The work of E.S. is funded by the Marie Curie Individual Fellowship MSCAIFEF-ST No. 660532-SuperMag. F.G, N.L. and A.F acknowledge the financial support of the European Research Council under the European Union’s Seventh Framework Program (FP7/2007-2013)/ERC Grant agreement No. 615187-COMANCHE and MIURFIRB2013 – Project Coca (Grant No. RBFR1379UX). I. M. acknowledges funding by the Academy of Finland grant No. 260880.

{\bf Methods.} Selenium-doped InAs NWs were grown by chemical beam epitaxy on a InAs 111B substrate. Gold catalyst colloid nanoparticles  with $40\,{\rm nm}$ of diameter  were dispersed on the  substrate  and InAs NWs were grown  at $400\,{\rm ^\circ C}$ for 60 mins using tertiarybutylarsine TBA ($1.5\,{\rm Torr}$), trimethylindium (TMI) ($0.6\,{\rm Torr}$) and tertiarybutylselenide (DtBSe) ($0.3\,{\rm Torr}$). Then  the growth temperature  was increased to $440\,{\rm ^\circ C}$  and the growth was proceeded for  another 50 mins. NWs typically had a diameter $d = 90\pm10\,{\rm nm}$ and length $\approx2.5\,{\rm \mu m}$. E-beam lithography has been performed using positive PMMA (AR-P 679.04), $20\,{\rm kV}$ acceleration voltage and a dose $320\,{\rm \mu C/cm^2}$. Development was done in a $1:3$ solution of AR 600-56 PMMA developer and isopropanol and possible residuals of the resist were removed with plasma oxygen cleaning. Just before the evaporation the samples were immersed for one minute in a $~48^{\circ}{\rm C} $ ${\rm (NH_4)_2S_x}$ solution to remove native oxide from to top of the NW to minimize unwanted scattering in the interface of the contacts. After this they were immersed for one minute in water and quickly rinsed in isopropanol before moving them into the ultra-high vacuum evaporator with a base vacuum of $10^{-10}\,{\rm Torr}$. The first $5\,{\rm nm}$-thick Ti layer was evaporated with $1\,{\rm \AA/s}$ rate to provide a sticking layer. $50\,{\rm nm}$-thick ${\rm Al}_{0.98}{\rm Mn}_{0.02}$ layer was then evaporated on top of the Ti with $1.5\,{\rm \AA/s}$ rate. Subsequently the samples were drawn back to the loading chamber for the oxidation in $0.2-0.4\,{\rm Torr}$ of ${\rm O_2}$ for $5$ minutes. The final evaporation was then completed again in the main chamber to obtain $50\,{\rm nm}$-thick pure Al layer with $~1.5\,{\rm A/s}$ rate. Lift-off was done in $~50^{\circ}{\rm C} $ acetone before finally rinsing the samples with isopropanol. For measuring, the samples were bonded with Al wires to 24-legged sample holder thus providing six NW devices to be measured with a single bonding. All the measurements were performed in a cryo-free $^3{\rm He}/^4{\rm He}$ dilution refrigerator with a base temperature of $30\,{\rm mK}$ . First the junctions were characterized applying voltage bias with a DC voltage source and measuring the current through room-temperature current preamplifier varying cryostat temperature and off-plane magnetic field. Filtering included two low-pass RC filters and two LC $\pi$-filters anchored at base temperature, and an additional LC $\pi$-filters stage at room temperature.

The theoretical comparison for the measured data was acquired using BSC theory for SINIS tunnel junctions assuming quasi-equilibrium of the electrons in the normal metal. The current $I$ through a SINIS junction as a function of bias voltage $V$ can be expressed as:

\begin{equation}
I(V) = \frac{1}{eR_N}\int^{\infty}_{-\infty}n_S\left[f_N(E-e\frac{V}{2})-f_N(E+e\frac{V}{2})\right]dE,
\label{Eq:IV}
\end{equation} 

\noindent where $R_N$ is the total normal-state resistance of the junction,

\begin{equation}
n_S = \left|{\rm Re}\left[\frac{E+i\gamma\Delta}{\sqrt{(E-i\gamma\Delta)^2-\Delta^2}}\right]\right|
\end{equation}

\noindent is the normalized density of states in the superconductor~\cite{giazotto2006opportunities} with Dynes parameter $\gamma$ measuring the life-time broadening of the quasi-particles or photon assisted tunneling leading to non-ideal behavior of the superconductor~\cite{dynes1978direct}. Also the temperature dependence of $\Delta(T)$ is taken into account as well. In the above expression,   

\begin{equation}
f_N(E) = \frac{1}{e^{E/k_BT_e}+1}
\end{equation}

\noindent is the Fermi--Dirac distribution of the normal metal. Conductance was obtained by differentiating Eq.~(\ref{Eq:IV}).


\begin{thebibliography}{10}
\newcommand{\enquote}[1]{``#1''}
\expandafter\ifx\csname urlstyle\endcsname\relax
  \providecommand{\doi}[1]{doi:\discretionary{}{}{}#1}\else
  \providecommand{\doi}{doi:\discretionary{}{}{}\begingroup
  \urlstyle{rm}\Url}\fi

\bibitem{giazotto2012josephson}
F.~Giazotto and M.~J. Mart{\'\i}nez-P{\'e}rez.
\newblock \enquote{The Josephson heat interferometer}.
\newblock \emph{Nature}, 492(7429), 401--405, 2012.

\bibitem{altimiras2010non}
C.~Altimiras, H.~Le~Sueur, U.~Gennser, et~al.
\newblock \enquote{Non-equilibrium edge-channel spectroscopy in the integer
  quantum Hall regime}.
\newblock \emph{Nature Physics}, 6(1), 34--39, 2010.

\bibitem{muhonen2012micrometre}
J.~T. Muhonen, M.~Meschke, and J.~P. Pekola.
\newblock \enquote{Micrometre-scale refrigerators}.
\newblock \emph{Reports on Progress in Physics}, 75(4), 046501, 2012.

\bibitem{giazotto2006opportunities}
F.~Giazotto, T.~T. Heikkil{\"a}, A.~Luukanen, et~al.
\newblock \enquote{Opportunities for mesoscopics in thermometry and
  refrigeration: Physics and applications}.
\newblock \emph{Reviews of Modern Physics}, 78(1), 217, 2006.

\bibitem{fornieri2016towards}
A.~Fornieri and F.~Giazotto.
\newblock \enquote{Towards phase-coherent caloritronics in superconducting
  quantum circuits}.
\newblock \emph{arXiv preprint arXiv:1610.01013}, 2016.

\bibitem{dresselhaus2007new}
M.~S. Dresselhaus, G.~Chen, M.~Y. Tang, et~al.
\newblock \enquote{New Directions for Low-Dimensional Thermoelectric
  Materials}.
\newblock \emph{Advanced Materials}, 19(8), 1043--1053, 2007.

\bibitem{wu2013large}
P.~M. Wu, J.~Gooth, X.~Zianni, et~al.
\newblock \enquote{Large thermoelectric power factor enhancement observed in
  InAs nanowires}.
\newblock \emph{Nano letters}, 13(9), 4080--4086, 2013.

\bibitem{vineis2010nanostructured}
C.~J. Vineis, A.~Shakouri, A.~Majumdar, et~al.
\newblock \enquote{Nanostructured thermoelectrics: big efficiency gains from
  small features}.
\newblock \emph{Advanced Materials}, 22(36), 3970--3980, 2010.

\bibitem{li2003thermal}
D.~Li, Y.~Wu, P.~Kim, et~al.
\newblock \enquote{Thermal conductivity of individual silicon nanowires}.
\newblock \emph{Applied Physics Letters}, 83(14), 2934--2936, 2003.

\bibitem{yazji2015complete}
S.~Yazji, E.~A. Hoffman, D.~Ercolani, et~al.
\newblock \enquote{Complete thermoelectric benchmarking of individual InSb
  nanowires using combined micro-Raman and electric transport analysis}.
\newblock \emph{Nano Research}, 8(12), 4048--4060, 2015.

\bibitem{roddaro2013giant}
S.~Roddaro, D.~Ercolani, M.~A. Safeen, et~al.
\newblock \enquote{Giant thermovoltage in single InAs nanowire field-effect
  transistors}.
\newblock \emph{Nano letters}, 13(8), 3638--3642, 2013.

\bibitem{tikhonov2016local}
E.~Tikhonov, D.~Shovkun, D.~Ercolani, et~al.
\newblock \enquote{Local noise in a diffusive conductor}.
\newblock \emph{Scientific Reports}, (6), 30621, 2016.

\bibitem{mourik2012signatures}
V.~Mourik, K.~Zuo, S.~M. Frolov, et~al.
\newblock \enquote{Signatures of Majorana fermions in hybrid
  superconductor-semiconductor nanowire devices}.
\newblock \emph{Science}, 336(6084), 1003--1007, 2012.

\bibitem{plissard2013formation}
S.~R. Plissard, I.~van Weperen, D.~Car, et~al.
\newblock \enquote{Formation and electronic properties of InSb nanocrosses}.
\newblock \emph{Nature nanotechnology}, 8(11), 859--864, 2013.

\bibitem{chang2013tunneling}
W.~Chang, V.~Manucharyan, T.~S. Jespersen, et~al.
\newblock \enquote{Tunneling spectroscopy of quasiparticle bound states in a
  spinful Josephson junction}.
\newblock \emph{Physical review letters}, 110(21), 217005, 2013.

\bibitem{larsen2015semiconductor}
T.~Larsen, K.~Petersson, F.~Kuemmeth, et~al.
\newblock \enquote{Semiconductor-nanowire-based superconducting qubit}.
\newblock \emph{Physical review letters}, 115(12), 127001, 2015.

\bibitem{miller2008high}
N.~Miller, G.~O’Neil, J.~Beall, et~al.
\newblock \enquote{High resolution x-ray transition-edge sensor cooled by
  tunnel junction refrigerators}.
\newblock \emph{Applied Physics Letters}, 92(16), 163501, 2008.

\bibitem{giazotto2008ultrasensitive}
F.~Giazotto, T.~T. Heikkil{\"a}, G.~P. Pepe, et~al.
\newblock \enquote{Ultrasensitive proximity Josephson sensor with kinetic
  inductance readout}.
\newblock \emph{Applied Physics Letters}, 92(16), 162507, 2008.

\bibitem{martinez2014quantum}
M.~J. Mart{\'\i}nez-P{\'e}rez and F.~Giazotto.
\newblock \enquote{A quantum diffractor for thermal flux}.
\newblock \emph{Nature communications}, 5, 2014.

\bibitem{fornieri2015nanoscale}
A.~Fornieri, C.~Blanc, R.~Bosisio, et~al.
\newblock \enquote{Nanoscale phase engineering of thermal transport with a
  Josephson heat modulator}.
\newblock \emph{Nature nanotechnology}, 2015.

\bibitem{martinez2015rectification}
M.~J. Mart{\'\i}nez-P{\'e}rez, A.~Fornieri, and F.~Giazotto.
\newblock \enquote{Rectification of electronic heat current by a hybrid thermal
  diode}.
\newblock \emph{Nature nanotechnology}, 10(4), 303--307, 2015.

\bibitem{leijnse2014thermoelectric}
M.~Leijnse.
\newblock \enquote{Thermoelectric signatures of a Majorana bound state coupled
  to a quantum dot}.
\newblock \emph{New Journal of Physics}, 16(1), 015029, 2014.

\bibitem{lopez2014thermoelectrical}
R.~L{\'o}pez, M.~Lee, L.~Serra, et~al.
\newblock \enquote{Thermoelectrical detection of Majorana states}.
\newblock \emph{Physical Review B}, 89(20), 205418, 2014.

\bibitem{pekola2004limitations}
J.~P. Pekola, T.~Heikkil{\"a}, A.~Savin, et~al.
\newblock \enquote{Limitations in cooling electrons using
  normal-metal-superconductor tunnel junctions}.
\newblock \emph{Physical review letters}, 92(5), 056804, 2004.

\bibitem{quaranta2011cooling}
O.~Quaranta, P.~Spathis, F.~Beltram, et~al.
\newblock \enquote{Cooling electrons from 1 to 0.4 K with V-based
  nanorefrigerators}.
\newblock \emph{Applied Physics Letters}, 98(3), 032501, 2011.

\bibitem{nevala2012sub}
M.~Nevala, S.~Chaudhuri, J.~Halkosaari, et~al.
\newblock \enquote{Sub-micron normal-metal/insulator/superconductor tunnel
  junction thermometer and cooler using Nb}.
\newblock \emph{Applied Physics Letters}, 101(11), 112601, 2012.

\bibitem{gunnarsson2015interfacial}
D.~Gunnarsson, J.~Richardson-Bullock, M.~J. Prest, et~al.
\newblock \enquote{Interfacial engineering of semiconductor--superconductor
  junctions for high performance micro-coolers}.
\newblock \emph{Scientific reports}, 5, 2015.

\bibitem{svensson2013nonlinear}
S.~F. Svensson, E.~A. Hoffmann, N.~Nakpathomkun, et~al.
\newblock \enquote{Nonlinear thermovoltage and thermocurrent in quantum dots}.
\newblock \emph{New Journal of Physics}, 15(10), 105011, 2013.

\bibitem{doh2005tunable}
Y.-J. Doh, J.~A. van Dam, A.~L. Roest, et~al.
\newblock \enquote{Tunable supercurrent through semiconductor nanowires}.
\newblock \emph{science}, 309(5732), 272--275, 2005.

\bibitem{roddaro2011hot}
S.~Roddaro, A.~Pescaglini, D.~Ercolani, et~al.
\newblock \enquote{Hot-electron effects in InAs nanowire Josephson junctions}.
\newblock \emph{Nano Research}, 4(3), 259--265, 2011.

\bibitem{bjork2002one}
M.~Bj{\"o}rk, B.~Ohlsson, T.~Sass, et~al.
\newblock \enquote{One-dimensional steeplechase for electrons realized}.
\newblock \emph{Nano Letters}, 2(2), 87--89, 2002.

\bibitem{roddaro2011manipulation}
S.~Roddaro, A.~Pescaglini, D.~Ercolani, et~al.
\newblock \enquote{Manipulation of electron orbitals in hard-wall InAs/InP
  nanowire quantum dots}.
\newblock \emph{Nano letters}, 11(4), 1695--1699, 2011.

\bibitem{romeo2012electrostatic}
L.~Romeo, S.~Roddaro, A.~Pitanti, et~al.
\newblock \enquote{Electrostatic spin control in InAs/InP nanowire quantum
  dots}.
\newblock \emph{Nano letters}, 12(9), 4490--4494, 2012.

\bibitem{rossella2014nanoscale}
F.~Rossella, A.~Bertoni, D.~Ercolani, et~al.
\newblock \enquote{Nanoscale spin rectifiers controlled by the Stark effect}.
\newblock \emph{Nature nanotechnology}, 9(12), 997--1001, 2014.

\bibitem{viti2012se}
L.~Viti, M.~S. Vitiello, D.~Ercolani, et~al.
\newblock \enquote{Se-doping dependence of the transport properties in
  CBE-grown InAs nanowire field effect transistors}.
\newblock \emph{Nanoscale research letters}, 7(1), 1--7, 2012.

\bibitem{suyatin2007sulfur}
D.~Suyatin, C.~Thelander, M.~Bj{\"o}rk, et~al.
\newblock \enquote{Sulfur passivation for ohmic contact formation to InAs
  nanowires}.
\newblock \emph{Nanotechnology}, 18(10), 105307, 2007.

\bibitem{ruggiero2004dilute}
S.~Ruggiero, A.~Williams, W.~Rippard, et~al.
\newblock \enquote{Dilute Al-Mn alloys for low-temperature device
  applications}.
\newblock \emph{Journal of low temperature physics}, 134(3-4), 973--984, 2004.

\bibitem{dynes1978direct}
R.~C. Dynes, V.~Narayanamurti, and J.~P. Garno.
\newblock \enquote{Direct Measurement of Quasiparticle-Lifetime Broadening in a
  Strong-Coupled Superconductor}.
\newblock \emph{Phys. Rev. Lett.}, 41, 1509--1512, 1978.
\newblock \doi{10.1103/PhysRevLett.41.1509}.

\end{thebibliography}
\end{document}